\documentstyle[12pt]{article}
\newcommand{\np}{Nucl. Phys.\ } 
\newcommand{\pl}{Phys. Lett.\ } 
\newcommand{\pr}{Phys. Rev.\ } 
 
\def\del{\partial} 
\def\Tr{\hbox{\rm Tr}} 
\def\cp{{c'\!}} 
\def\hat{\widehat} 
\def\tilde{\widetilde} 
\def\OSW{{\cal O}_{SW}} 
\def\qbar{{\overline{q}}} 
\def\slash#1{\mbox{$\not \!\! #1$}} 
\def\Dslash{{\slash D}} 
\def\rDslash{{\overrightarrow{\slash D}}} 
\def\lDslash{{\overleftarrow{\slash D}}} 
\newcommand{\dslash}{\not \hspace{-2.5pt} \partial} 
\newcommand{\ldslash}{{\overleftarrow{\dslash}}} 
\newcommand{\pslash}{\not \hspace{-3pt} p} 
\def\BRST{{\scriptstyle\rm BRST}} 
\def\NGI{{\scriptstyle\rm NGI}}

\def\amp{\mbox{\scriptsize amp}}
 
\def\spose#1{\hbox to 0pt{#1\hss}} 
\def\ltapprox{\mathrel{\spose{\lower 3pt\hbox{$\mathchar"218$}} 
 \raise 2.0pt\hbox{$\mathchar"13C$}}} 
\def\gtapprox{\mathrel{\spose{\lower 3pt\hbox{$\mathchar"218$}} 
 \raise 2.0pt\hbox{$\mathchar"13E$}}} 
\def\inapprox{\mathrel{\spose{\lower 3pt\hbox{$\mathchar"218$}} 
 \raise 2.0pt\hbox{$\mathchar"232$}}} 
 
\begin{document} 
 
\pagestyle{empty} 
\begin{flushright} 
CERN-TH/2001-100\\ 
ROME1-1314/2001 \\ 
ROM2F/2001/13\\ 
SHEP 01/13 \\ 
UW/PT 01-13 \\ 
\end{flushright} 

\centerline{\LARGE{\bf Non-perturbative improvement}} 
\vskip 0.2cm 
\centerline{\LARGE{\bf of lattice QCD at large momenta}} 
\vskip 0.3cm 
\vskip 1cm 
\centerline{\bf{G.~Martinelli$^a$, G.C.~Rossi$^b$~\footnote{On leave 
of absence from Dipartimento di Fisica, Universit\'a di Roma 
``{\it Tor Vergata}'', Via della Ricerca Scientifica, 
I-00133 Roma, Italy}, 
C.T.~Sachrajda$^c$, S.~Sharpe$^{d}$,}} 
\vskip 0.2cm 
\centerline{\bf{M.~Talevi$^a$ and M.~Testa$^{a}$}} 
\vskip 0.3cm 
\centerline{$^a$ Dip. di Fisica, Univ. di Roma ``La Sapienza'' and 
INFN, Sezione di Roma ``La Sapienza''} 
\centerline{P.le A. Moro 2, I-00185 Roma, Italy} 
\smallskip 
\centerline{$^b$ Theory Division, CERN, 1211 Geneva 23, Switzerland.} 
\smallskip 
\centerline{$^c$ Dept. of Physics and Astronomy, 
University of Southampton} 
\centerline{Southampton SO17 1BJ, UK} 
\smallskip 
\centerline{$^d$ Physics Department, University of Washington, 
Seattle WA 98195, USA} 
\vskip 2cm 
\centerline{\bf ABSTRACT} 
\begin{quote} 
{We propose a method to improve lattice operators composed of Wilson fermions 
which allows the removal of all corrections of $O(a)$, including those 
proportional to the quark mass. It requires off-shell improvement of quark 
fields and composite operators, which is achieved by studying the behaviour 
of quark and gluon correlation functions at large momenta.} 
\end{quote} 
\vfill 
\newpage 
 
\pagestyle{plain} 
\setcounter{page}{1} 
 
\section{Introduction} 
\label{sec:introd} In present lattice computations an important 
source of errors is the finiteness of the lattice spacing, $a$. A 
systematic method for reducing discretization errors order by 
order in $a$ was proposed by Symanzik~\cite{Symanzik} and 
developed by L\"uscher and Weisz~\cite{Luscher}. It consists of 
modifying the action and operators by ``irrelevant'' terms, chosen 
in such a way that the convergence to the continuum is 
accelerated. In the first implementations of this procedure, the 
improvement coefficients were computed in perturbation theory, 
leaving errors of O$(a g_0^{2})$ in physical 
quantities~\cite{imppert}. It turns out, however, that, for 
lattice spacings used in present simulations, one-loop 
perturbation theory is insufficiently precise for some 
quantities~\cite{bknonpert,jlqcd}, even when it is ``tadpole 
improved''~\cite{lepagemack}. In refs.~\cite{Luscher2,Luscher3} a 
method for determining the improvement coefficients beyond 
perturbation theory, based on the enforcement of Ward-Takahashi 
identities (WTI's) has been proposed and implemented, thus 
achieving full $O(a^2)$ improvement. This method allows one to 
determine the improved action and vector current for massive 
quarks (see also ref.~\cite{guasom}), but can only be applied to 
other operators, including the axial current, in the chiral limit. 
The reason for this limitation and a strategy to work outside the 
chiral limit valid in the case of on-shell matrix elements was 
discussed in ref.~\cite{mrsstt}. Vector and axial WTI's, possibly 
in conjunction with the  use of unequal quark masses, have been 
employed to determine the  coefficient of the term proportional to 
the mass in the normalization constants of certain quark 
bilinears~\cite{gmdd,bcgls}. The question of O($a m_q$) 
corrections~\footnote{By $m_q$ we mean some choice of physical 
quark mass. We do not need to pick a specific definition in this 
paper.} is particularly significant for applications involving 
heavy quarks, where these terms may be large. Important examples 
in the phenomenology of $B$-mesons include the calculations of the 
leptonic decay constant $f_B$, the form factors of semileptonic 
$B$ decays, the $B$-parameters and the amplitudes of the radiative 
decay  $B\to K^*\gamma$. 
 
In this paper we suggest an alternative method to extend operator 
improvement beyond the chiral limit, based on the consideration of 
the high momentum behaviour of quark correlators. The basic 
ingredient is the matching of lattice and continuum Green 
functions at large Euclidean momenta (or equivalently at short 
distances), where perturbation theory applies. An important 
property, crucial for our purposes, is that renormalized continuum 
correlators do not contain contact terms beyond those found in 
perturbation theory. This is a  general consequence of the 
Callan-Symanzik equation, when applied at large momenta in an 
asymptotically free field theory. We stress that we aim only to 
remove errors proportional to $a$ and do not attempt to work at 
arbitrary values of $a m_q$, as in the program proposed in 
ref.~\cite{FNAL}. In particular, after implementation of our 
proposal there will still remain errors proportional to $a^2 
m_q^2$.  
 
Our method allows for the evaluation of the 
renormalization constants of divergent operators, such as the 
scalar and pseudoscalar densities. For a discussion of recent 
determinations of $Z_P$, $Z_S$ and their ratio see 
ref.~\cite{LUB}. In principle this approach can be 
applied to composite operators of arbitrary dimension, though at 
the price of a very high proliferation of mixing operators, which 
may render the actual implementation of this scheme impractical. 
 
We wish to mention, as a side remark, an important limitation of 
the general idea of exploiting the large-momentum behaviour of 
correlators to renormalize composite operators non-perturbatively. 
The point is that for composite operators which can mix with lower 
dimensional ones, such as the octet part of the weak Hamiltonian, 
the strategy of killing at large momenta the ``form factors'' not 
allowed by chiral symmetry cannot be made to work. In fact lower 
dimensional operators may give rise to a large-momentum behaviour 
that cannot be distinguished from that due to the exchange of 
Goldstone bosons. For completeness we will briefly discuss this 
question in Appendix A. 
 
An alternative method~\cite{dawson}, which does not suffer from 
these limitations, is to measure on the lattice the behaviour of 
correlation functions in $x$-space in the region $a\ll |x| \ll 
\Lambda_{QCD}^{-1}$, where perturbation theory is reliable, and 
use the O.P.E. with perturbatively computed Wilson coefficients to 
fit this dependence. The parameters determined by the fitting 
procedure are the physical matrix elements of the (finite, 
renormalized) operators that appear in the O.P.E. In this way 
there is no need to know the expression of the renormalized 
operators in terms of the bare operators of the regularized 
(lattice) theory. The question here is whether the lattice is 
sufficiently fine-grained for the window $a\ll |x| \ll 
\Lambda_{QCD}^{-1}$ to exist. If this is the case, the method can 
be applied to a number of interesting cases, some of which are 
discussed in refs.~\cite{dawson} and~\cite{marti}. 
 
The approach presented in this paper is a natural extension, valid 
up to O($a^2$), of the method proposed in ref.~\cite{MPSTV}, where 
the renormalization procedure is carried out in a non-perturbative 
way on correlators with external quarks and gluons. As discussed 
in ref.~\cite{MPSTV}, such a procedure is expected to work if we 
can choose the virtuality of the external states, $\mu$, at which 
the normalization is carried out, so as to satisfy $\Lambda_{\rm 
QCD} \ll \mu \ll 1/a$. Without improvement, the relative errors in 
matrix elements are of order $a m_q$ and $a \Lambda_{\rm QCD}$. In 
principle, our method is able to remove these errors and leave 
only corrections of O($a^2$). 
 
As has been said above, the condition $ \Lambda_{\rm QCD} \ll \mu$ 
is required so that one can use perturbation theory to relate 
matrix elements renormalized on the lattice to those computed in 
standard continuum schemes. We wish to stress that, in principle, 
we can relax this constraint by using a sequence of lattices of 
decreasing $a$ and correspondingly decreasing physical volumes. 
The idea is to determine the normalization constants by imposing 
renormalization conditions on a lattice with a very small lattice 
spacing. On such a fine-grained lattice large values of $\mu$ are 
allowed, but the smallness of the physical volume precludes the 
possibility of reliably computing hadronic correlation functions. 
It is therefore necessary to determine the corresponding 
normalization constants for the larger, but coarser, lattice on 
which the physical matrix elements are finally computed. This is 
achieved by increasing the lattice spacing $a$ and matching the 
normalization constants at fixed $\mu$, and successively 
decreasing $\mu$ and matching at fixed $a$. The procedure is 
repeated until the required value of $a$ is reached. This method, 
which has been proposed within the context of the Schr\"odinger 
functional in ref.~\cite{Luscher2}, could be applied also here. A 
first attempt in this direction has been made 
recently~\cite{Zhestov}. 
 
Our approach requires fixing the gauge, a procedure usually 
afflicted by Gribov ambiguities~\cite{GRIBOV}~\footnote{For recent 
progress on this issue see refs.~\cite{NEUB,MAX}.}. However, in 
the large-momentum region, where correlators are well described by 
perturbation theory, these ambiguities vanish asymptotically. 
Gauge fixing does however, introduce a further complication. When 
we evaluate correlators of gauge invariant operators with external 
quark and gluon fields, we need, in general, to include mixing 
with gauge non-invariant operators. The form of the latter is 
restricted by BRST symmetry~\cite{zuber,BBH}. For flavour 
non-singlet quark bilinears, these restrictions are sufficient to 
forbid the appearance of gauge non-invariant terms until O($a^2$). 
For higher dimension operators, such as the four-fermion operators 
in the effective weak Hamiltonian, gauge non-invariant operators 
should be considered even at O($a^0$). The investigation of the 
consequences of BRST symmetry for the evaluation of Green 
functions of gauge-invariant operators between off-shell quark and 
gluon states is one of the principal goals of this paper. Our 
conclusions and the implications for the improvement of lattice 
actions and operators are presented in Sec.~\ref{sec:strate} 
below. 
 
The paper is organized as follows. In Sec.~\ref{sec:strate} we 
explain our strategy and apply it to the improvement of quark 
fields. In Sec.~\ref{sec:bilinears} we discuss the procedure to 
improve quark bilinears fully up to O($a^2$). Few conclusive 
comments can be found in Sec.~\ref{sec:conclusions}. The paper 
ends with three Appendices. In Appendix~A we discuss in a simple 
example why in presence of spontaneous chiral symmetry breaking it 
is not possible to renormalize non-perturbatively operators that 
can mix with lower dimensional operators by only looking at the 
high momentum behaviour of correlators. In Appendix~B we give a 
simple argument to prove that in off-shell amplitudes 
BRST-symmetry does not forbid the mixing of gauge-invariant 
operators with gauge non-invariant ones, if the latter vanish by 
the equations of motion. In Appendix~C we apply our off-shell 
improvement procedure to the quark propagator calculated at 
one-loop order in perturbation theory. 
 
\section{Strategy and implementation on bilinears} 
\label{sec:strate} 
 
Following ref.~\cite{Luscher2}, we recall the basic steps of the Symanzik 
improvement program for lattice QCD up to O($a^2$). We will refer to the 
standard Wilson formulation of the theory. The improvement of the spectrum 
can be accomplished by adding the SW-Clover operator to the lattice QCD 
action with a coefficient determined using suitable WTI's identities for the 
axial current. This coefficient is a function of $g_0^2$, and has been 
determined non-perturbatively for several values of $g_0^2$ using numerical 
simulations~\cite{Luscher3}. The improvement of the action is not sufficient 
to remove O($a$) terms from correlation functions~\cite{Luscher,imppert}. To 
achieve this result one must also improve the relevant operators, by adding 
to them terms of three kinds~\footnote{This classification was introduced in 
the study of renormalization of gauge invariant operators, which requires, in 
general, operators of all three types~\cite{zuber}~\cite{BBH}. The 
classification applies as well for improvement, since the allowed operators 
are restricted by the same type of symmetries relevant for mixing.}: 
\begin{itemize} \item[(a)] Operators which vanish by the equations of 
motion,  which can be either gauge invariant or non-invariant.  They give 
rise only to contact terms,  and thus do not affect the long-distance 
properties of correlators. \item[(b)] Gauge non-invariant, but BRST 
invariant, operators, which also do not contribute to physical amplitudes. 
\item[(c)] 
Gauge-invariant operators which do not vanish by the equations of motion. 
These are needed to cancel discretization errors proportional to $a$ 
in physical amplitudes. 
\end{itemize} 
Within the approach of ref.~\cite{Luscher2}, by working in the chiral limit, 
one is able to improve operators by  using  only physical matrix elements. 
Improvement of composite operators in on-shell matrix elements was extended 
beyond the chiral limit in ref.~\cite{mrsstt} (see also 
refs.~\cite{gmdd,bcgls}). In both cases only  terms of type (c) are 
required. Here we need to consider all three types of terms. 
 
The idea developed in this paper is to construct finite operators 
with well defined chiral transformation properties by matching 
lattice and continuum correlators in the region of large Euclidean 
momenta where perturbation theory applies. This is most easily 
done by requiring the vanishing of chirality-violating form 
factors in correlation functions in which the operator is inserted 
together with external quark and/or gluon (and possibly ghost) 
legs with large virtualities. This procedure is justified by the 
following two observations.  First, at large momenta renormalized 
perturbation theory becomes chirally invariant (note that: i) 
explicit chiral symmetry breaking effects induced by the 
regularization can be reabsorbed by imposing the validity of the 
WTI's of chiral symmetry and ii) violations from the non-vanishing 
of quark masses disappear at large momenta). Secondly, 
contributions due to the spontaneous breaking of chiral symmetry, 
which are absent in perturbation theory, die off at large momenta. 
So the magnitude of both effects (which come from both 
perturbative and non-perturbative violations of chirality) 
decreases as we go deeper into the Euclidean region. As discussed 
in Appendix A, the argument cannot be applied to determine the 
mixing coefficients of lower dimensional operators. 
 
In the following we consider correlation functions of elementary 
fields and multiplicatively renormalizable operators. They have, 
in general, non-vanishing anomalous dimensions. To discuss the 
improvement of such operators in a systematic way, one must first 
multiply them by appropriate renormalization constants so that 
they have a finite continuum limit. One can then speed up the 
approach to the continuum limit by adding further operators of 
O($a$), and by allowing the normalization constants of fields and 
operators to depend on $a m_q$. Thus, to the order at which we 
work, all the $Z$-factors we introduce will depend linearly on $a 
m_q$. We do not, however, exhibit this dependence explicitly -- it 
is fixed automatically by the overall normalization conditions 
that will be imposed. 
 
\subsection{Lattice BRST symmetry} 
 
In order to classify operators of type (b), i.e. gauge non-invariant 
operators which are BRST invariant, we need to know the form of the lattice 
BRST transformation. This is a simple generalization of the continuum 
transformation, as explained in ref.~\cite{LuscherBRST}. 
 
The lattice Landau gauge can be enforced by requiring that 
\begin{equation} 
G(U) = \sum_{n,\mu} \Tr(U_{n,\mu}+U_{n,\mu}^{\dagger}) 
\end{equation} 
be locally maximized along a gauge orbit. This is equivalent to 
imposing, at each site $n$, and for each color $a$, 
\begin{equation} 
0 = f_n^a(U) = {\delta G(U^g) \over \delta\omega_n^a}\bigg|_{\omega=0} 
 = i \sum_\mu \Tr\left[ t^a (U_{n,\mu} - U_{n,\mu}^{\dagger}) - 
t^a (U_{n-\mu,\mu} - U_{n-\mu,\mu}^{\dagger}) \right] \,, 
\label{eq:latticelandau} 
\end{equation} 
where $\omega_n^a$ parameterizes the gauge transformation at site 
$n$ and $g_n = \exp(i\omega_n^a t^a)$. The 
condition~(\ref{eq:latticelandau}) is the discretized version of 
$\partial_\mu A_\mu^a(x)=0$. We can implement the lattice Landau 
gauge in the functional integral in the standard way by adding to 
the action gauge fixing and ghost terms of the form 
\begin{equation} 
{\cal{S}}_{gf} = a^4\sum_n \left[ 
{1\over2}\alpha \lambda_n^a \lambda_n^a 
+ i \lambda_n^a f_n^a(U) 
+ \sum_{n'} \bar{c}_n^a 
{\delta f_n^a(U^g) \over \delta\omega_{n'}^b}\bigg|_{\omega=0} 
 c_{n'}^b 
\right] \,. \label{eq:gaugefix} 
\end{equation} 
In eq.~(\ref{eq:gaugefix}) $c$ and $\bar{c}$ are ghost and 
anti-ghost fields, while $\lambda$ is a Lagrange multiplier, all 
of which are defined on lattice sites. We obtain the lattice 
Landau gauge in the limit $\alpha\to0$, for then the integral over 
$\lambda_n^a$ sets $f_n^a=0$. 
 
There are, in general, many solutions to the gauge condition 
Eq.~(\ref{eq:latticelandau}): this is the problem of Gribov 
copies~\cite{GRIBOV,GRIBOVLAT}. As mentioned in the introduction, 
we assume here the absence of copies for the gauge configurations 
which are responsible for the dominant contributions to 
correlation functions at large momenta. 
 
The lattice action, ${\cal{S}}$, and the gauge-fixing action, 
${\cal{S}}_{gf}$, are separately invariant under the appropriate 
BRST transformations. The latter are the lattice generalization of 
the continuum transformations. They are constructed by letting 
gauge and fermion fields be gauge transformed with a gauge matrix 
$g_n(c) = \exp(i \epsilon c_n^a t^a)$, where $\epsilon$ is a 
constant Grassmann parameter: 
\begin{eqnarray} 
\epsilon\; \delta_{\rm BRST}\; q_n &=& g_n(c) q_n - q_n = 
 i \epsilon c_n^a t^a q_n 
\label{eq:quarkBRST}\\ 
\epsilon\; \delta_{\rm BRST}\; \qbar_n &=& - i \qbar_n \epsilon c_n^a t^a 
\label{eq:qbarBRST}\\ 
\epsilon\; \delta_{\rm BRST} \;U_{n,\mu} &=& 
i \epsilon \left(c_n^a t^a U_{n,\mu}- U_{n,\mu} t^a c_{n+\mu}^a\right) 
\label{eq:UBRST}\,. 
\end{eqnarray} 
Note that higher order terms vanish since $\epsilon^2=0$. 
The other fields transform in the usual way, i.e. 
\begin{equation} 
\delta_{\rm BRST}\; c_n^a = - {1\over2} f_{abc} c_n^b c_n^c \,,\quad 
\delta_{\rm BRST}\; \bar{c}_n^a = \lambda_n^a \,, \quad 
\delta_{\rm BRST}\; \lambda_n^a = 0 \,. 
\label{eq:cBRST} 
\end{equation} 
The lattice BRST transformation is nilpotent, like its continuum counterpart. 
 
The full action is also invariant under the anti-ghost shift 
symmetry, $\bar{c}_n^a \to \bar{c}_n^a + {\rm const}$. This 
symmetry is also very useful in restricting the form of improved 
operators. 
 
\subsection{Improvement of the action} 
 
The Wilson fermion action is improved by the addition of the SW operator, 
\begin{eqnarray} 
{\cal{S}} &=& {\cal{S}}_{\rm gauge} + {\cal{S}}_{\rm Wilson} + 
a \int d^4x\ \OSW 
\label{eq:swclover}\\ 
&\equiv& {\cal{S}}_{\rm gauge} + \int d^4x\ \qbar (\rDslash + m_0) q \,, 
\label{eq:dslashdef} 
\end{eqnarray} 
where 
\begin{equation} 
\OSW = - \frac{i}4 c_{SW}\ \sum_{\mu\nu} \qbar \sigma_{\mu\nu} F_{\mu\nu} q 
\,, 
\end{equation} 
with $\sigma_{\mu\nu} = (i/2)[\gamma_\mu,\gamma_\nu]$. In 
Eq.~(\ref{eq:dslashdef}), $\rDslash +m_0$ is a shorthand for the 
entire lattice  fermion operator. Here and in the following we use 
continuum notation to refer to lattice quantities. For physical 
quantities to approach their continuum values with deviations of 
only O($a^2$), we need, in addition to choosing the correct value 
of $c_{SW}$, to adjust the bare gauge coupling $g_0$ and the bare 
mass $m_0$ appropriately, in a way which depends on the 
renormalized quark mass. This is discussed in detail in 
ref.~\cite{Luscher2}, but will not concern us here, since we have 
in mind fixing $a(g_0)$ and $m_0$ using hadronic quantities, such 
as $m_\rho$ and $m_\pi$. 
 
There are two additional gauge invariant operators of dimension 5 
which could, in principle, be added to the action. The first one 
is 
\begin{eqnarray} 
{\cal{S}}_1' =  a m_0 \cp_{1} \int d^4x \, \qbar (\rDslash+m_0) q\, , 
\label{eq:c1pdef} 
\end{eqnarray} 
but this can be immediately eliminated by rescaling the quark fields. 
The second one can be written in the form 
\begin{eqnarray} 
{\cal{S}}_2' = 2 a \cp_2 \int d^4x \, \qbar (\rDslash + m_0)^2 q 
\label{eq:c2pdef} 
\end{eqnarray} 
and thus vanishes by the equations of motion. Although it can be neglected 
for the computation of on-shell quantities, it gives rise to contact terms 
in correlators. Its contributions can, however, be reabsorbed by a suitable 
redefinition of the O($a$) part of fundamental fields and operators, as 
discussed below. Consequently we do not include it in the action. 
 
Finally, we must consider the possibility that gauge non-invariant operators 
may need to be added to the action in order to improve quark and gluon 
correlation functions. Any such operator should be invariant under both the 
lattice BRST symmetry and the anti-ghost shift symmetry, and also under global 
color transformations, lattice rotations and translations. In addition, it 
must have zero ghost number. Allowed terms thus have the form (again using 
continuum notation for lattice quantities) 
\begin{equation} 
{\cal{S}}_{GNI} = \int d^4x\, \delta_{\rm BRST} 
\left[ \bar{c}^a  \partial_\mu X_\mu^a \right] 
\,, \quad 
X_\mu^a = X_\mu^a(\partial_\nu\bar{c},c,A,\bar{q},q,\lambda) 
\,, 
\end{equation} 
where the vector $X_\mu^a$ is a color octet with zero ghost 
number. The possibility of lowest dimension is $X_\mu^a=A_\mu^a$, 
with some choice for the lattice gauge field, $A_\mu^a$. This 
leads, however, to a term proportional to 
${\cal{S}}_{gf}(\alpha=0) + \mbox{O}(a^2)$, which can be absorbed 
by rescaling $\lambda$ and $\alpha$. Improvement terms of O($a$) 
would result from choices of an $X_\mu^a$ of dimension 2. There 
exist, however, no such operators, and therefore no possible gauge 
non-invariant improvement terms at O($a$). At next order, by 
contrast, there exist a number of possibilities, e.g. 
$X^a_\mu=\qbar \gamma_\mu t^a q$ or $X_\mu^a=f_{abc}{c}^b 
\partial_\mu\bar{c}^c$. 
 
\subsection{Improvement of quark fields} 
\label{sec:impquark} 
 
To improve quark fields to O($a^2$) we must add all possible 
operators of dimension $5/2$ which are allowed by the unbroken 
lattice symmetries. In particular, these operators must have the 
same properties as the quark fields under global gauge 
transformations, rotations and flavour transformations, and must 
satisfy the same BRST identities. The bare operators having these 
properties are $\Dslash q$, $m_0 q$ and $\dslash q$. The 
appearance of $\dslash q$ is, at first sight, surprising since it 
transforms differently from the quark field $q$ under local gauge 
transformations [and thus under BRST transformations, 
Eq.~(\ref{eq:quarkBRST})]. Nevertheless, as we now explain, the 
non-linearity of BRST symmetry does not exclude such an operator. 
 
The constraints which follow from BRST symmetry can be obtained by 
requiring the improved operators to satisfy (up to O($a^2$)) the 
same identities that are satisfied by the continuum operators. The 
continuum identities take the form 
\begin{eqnarray} 
0 &=& 
\langle \delta_{\rm BRST}\left[O_1(x_1) O_2(x_2) \dots 
\right] \rangle \nonumber \\ 
&=& \langle \delta_{\rm B} [O_1(x_1)] O_2(x_2) \dots \rangle 
\pm \langle O_1(x_1) \delta_{\rm B} [O_2(x_2)] \dots \rangle 
+ \dots \, , 
\label{eq:BRSTTID} 
\end{eqnarray} 
where $O_i$ are local composite operators located at arbitrary positions. The 
choice of sign on the second line depends on whether $O_i$ is bosonic or 
fermionic. From now on to lighten the notation we will use the abbreviation 
$\delta_{\rm B}$ for $\delta_\BRST$. 
Note that for the identities to be non-trivial the product of operators 
being varied must have ghost number $-1$. The conditions on the lattice 
operators become 
\begin{equation} 
\langle \widehat{[\delta_{\rm B} O_1]} \hat O_2 \dots \rangle 
\pm \langle \hat O_1 \widehat{[\delta_{\rm B} O_2]} \dots \rangle 
+ \dots  = \mbox{O}(a^2) \,, 
\label{eq:BRSTIDimp} 
\end{equation} 
where $\hat O_i$ and $\widehat{[\delta_{\rm B} O_i]}$ are, respectively, 
the improved lattice versions of the continuum operator and of its 
continuum BRST variation. For brevity, we have subsumed the site label into 
the definition of the operator. The expectation value is to be taken 
with respect to the improved lattice action. 
 
Since the improvement of the action does not 
break the lattice BRST symmetry, 
one can derive analogous BRST identities directly on the lattice 
\begin{equation} 
\langle \delta_{\rm B}[\hat O_1] \hat O_2 \dots \rangle 
\pm \langle \hat O_1 \delta_{\rm B}[\hat O_2] \dots \rangle 
+ \dots  = 0 \,, 
\label{eq:BRSTIDlat} 
\end{equation} 
where the variations are under the lattice 
transformations~(\ref{eq:quarkBRST})--(\ref{eq:cBRST}). 
These identities are of the required form~(\ref{eq:BRSTIDimp}) if 
\begin{equation} 
\widehat{[\delta_{\rm B} O_i]} = 
\delta_{\rm B} [\hat O_i] + \mbox{O}(a^2) \,. 
\label{eq:BRSTfinal} 
\end{equation} 
In other words, the improved lattice version of the continuum BRST variation 
on the l.h.s. must equal (up to O($a^2$)) the lattice BRST variation of the 
improved operator on the r.h.s. 
 
We now discuss the consequences of the condition~(\ref{eq:BRSTfinal}). Since 
lattice and continuum BRST symmetries take the same form, 
(\ref{eq:BRSTfinal}) is automatically satisfied up to O($a$). 
Eq.~(\ref{eq:BRSTfinal}) is thus a constraint on the form of improvement 
terms. 
 
Consider first a gauge-invariant operator composed of quark and gluon fields. 
For this the continuum BRST variation vanishes and the 
condition~(\ref{eq:BRSTfinal}) 
simply becomes $\delta_{\rm B} [\hat O_i] = \mbox{O}(a^2)$. In other words, 
the terms of O($a$) added to improve $O_i$ should themselves be invariant 
under the lattice BRST transformation, with the exception of operators 
which vanish by the equations of motion. These do not need to be BRST 
invariant because they contribute only when the insertion points of two 
operators happen to coincide, in which case the 
constraint~(\ref{eq:BRSTfinal}) must be applied to the resulting composite 
operator. We do not discuss this issue further since, as explained below, 
BRST non-invariant operators are not needed for O($a^2$) improvement. 
 
For the case of a quark field the continuum BRST variation is non-vanishing 
and the condition~(\ref{eq:BRSTfinal}) becomes 
\begin{equation} 
i t^a \widehat{ [c^a q]} = \delta_{\rm B} [\widehat{q}] + O(a^2)\,. 
\label{eq:BRSTforq} 
\end{equation} 
As noted above, symmetries other than BRST require that the 
improved quark field has the form 
\begin{equation} 
\widehat q = Z_q^{-1/2} \left[1 +a \cp_q(\rDslash + m_0) 
+ a c_\NGI\dslash \right]q \,. 
\label{eq:imprquark} 
\end{equation} 
Here $Z_q$ contains an implicit dependence on the quark mass, and we have 
chosen to group $\rDslash$ with $m_0$ since this combination gives only 
contact terms in correlation functions. From Eq.~(\ref{eq:quarkBRST}) the 
lattice BRST variation of the improved quark field is 
\begin{equation} 
\delta_{\rm B} [\widehat{q}] = Z_q^{-1/2} i t^a \left\{ 
c^a \left[1 +a \cp_q(\rDslash + m_0)\right] q + 
a c_\NGI\dslash[c^a q] \right\} \,. 
\label{eq:dBRSTq} 
\end{equation} 
Thus, comparing~(\ref{eq:dBRSTq}) with~(\ref{eq:BRSTforq}) 
and using~(\ref{eq:imprquark}), we learn that 
\begin{equation} 
\widehat{ [c^a q]} = c^a\widehat{q} + 
a Z_q^{-1/2} c_\NGI [\dslash c^a] q \,. 
\end{equation} 
Naively, we might have expected the second term to be absent, 
since the ghost field itself does not require improvement at O($a$). 
But in fact we have no {\em a priori} knowledge of how to improve the 
composite operator $c^a q$, and there is no inconsistency with having the 
additional term proportional to $c_\NGI$. We conclude that BRST symmetry does 
not forbid non gauge invariant improvement terms for the quark field. For 
completeness we note that the improved antiquark field takes the form 
\begin{equation} \widehat\qbar = Z_q^{-1/2}\,\qbar 
\left[1 + a \cp_q(-\lDslash + m_0) - a c_\NGI\ldslash \right] \,. 
\label{eq:impqbar} 
\end{equation} 
 
Having determined the general form of the improved quark and antiquark fields, 
we now explain our method for determining non-perturbatively the improvement 
coefficients $\cp_q$ and $c_\NGI$ and the normalization factor, $Z_q$. 
 
To the order in $a$ at which we are working, $Z_q$ has a linear dependence 
on the quark mass, while $\cp_q$ and $c_\NGI$ are mass independent. 
The improvement condition  is that the lattice and continuum renormalized 
quark propagators,  $\hat{S}$ and $S$ respectively, differ only at 
O($a^2$),~i.e. 
\begin{equation} 
\hat{S}(p) =  S(p) + \mbox{O}(a^2)\,. 
\label{eq:impcond} 
\end{equation} 
Here, the renormalized lattice quark propagator in momentum space is defined by 
\begin{equation} 
\hat{S}(p) \equiv \int d^4x e^{-i p\cdot x} \hat{S}(x) \,, 
\end{equation} 
where 
\begin{eqnarray} 
&\!&\hat{S}(x) \equiv \langle \hat q(x) \hat\qbar(0) \rangle \\ 
&\!& = Z_q^{-1} \left\langle \left[1 +a \cp_q(\rDslash + m_0) + 
a c_\NGI \dslash \right] q(0) \qbar(x)\left[1 + a \cp_q(-\lDslash + m_0)- 
a c_\NGI \ldslash \right] \right\rangle \,.\nonumber 
\end{eqnarray} 
In the following, we denote the bare lattice quark propagator by $S_L$ 
($S_L(x) = \langle q(x) \qbar(0) \rangle$). 
 
The continuum renormalized quark propagator has the decomposition 
\begin{equation} 
S(p) =  i \Sigma_1(p^2) \pslash + \Sigma_2(p^2) 
\label{eq:decomposition} 
\end{equation} 
with some choice of normalization condition, to be specified 
below. The large $p^2$ behaviour of $\Sigma_1$ and $\Sigma_2$ is 
determined by the renormalization group equation~\cite{Politzer}. 
In particular, $\Sigma_1(p^2) 
\stackrel{p^2\to\infty}{\longrightarrow} 1/p^2$ up to computable 
logarithmic corrections, while, in the chiral limit, 
$\Sigma_2(p^2) \stackrel{p^2\to\infty}{\longrightarrow} 
\langle\qbar q\rangle/p^4$, again up to logarithms, where 
$\langle\qbar q\rangle$ is the usual order parameter for 
spontaneous chiral symmetry breaking. Away from the chiral limit, 
$\Sigma_2$ still vanishes at large $p^2$,  but now decreases as 
$m_q/p^2$. As already indicated in Eq.~(\ref{eq:impcond}), our 
improvement condition will be that the lattice propagator has the 
same asymptotic behaviour as the continuum one up to O($a^2$). 
 
The effect of the improvement terms can be seen by deriving the 
expression for the improved propagator in momentum space. One gets 
\begin{eqnarray} 
\hat{S}(p) &=& Z_q^{-1} \left(S_L(p) + 2 a \cp_q 
+ a c_\NGI \{i\pslash,S_L(p)\} + \mbox{O}(a^2) \right) 
\label{eq:impS}\\ 
&=& Z_q^{-1} \left(S_L(p) + 2 a \cp_q \right) 
+ 2 a c_\NGI \left[2 i\pslash \Sigma_2(p^2)- p^2\Sigma_1(p^2)\right] 
+ \mbox{O}(a^2) \,. 
\end{eqnarray} 
In the second line we have used Eq.~(\ref{eq:decomposition}) to 
parametrize the form of the propagator. The term proportional to 
$\cp_q$ thus adds a constant to the propagator, corresponding to a 
delta-function contact term in position space. The contribution 
from the term proportional to $c_\NGI$ is more complicated. In the 
chiral limit, and for large $p^2$, we can ignore $\Sigma_2$, and 
the $c_\NGI$ term is proportional to $p^2\Sigma_1(p^2)$. As 
already noted, this combination varies only logarithmically with 
$p^2$. Thus, in the chiral limit, the effect of the two 
improvement terms is similar. In practice  this leads to problems 
in separately determining the two coefficients $\cp_q$ and 
$c_\NGI$~\cite{LUB}. Away from the chiral limit, the $c_\NGI$ term 
also contributes a mass-dependent renormalization of the 
coefficient of $\pslash$. 
 
	From this analysis we thus expect that the unimproved lattice 
propagator does not decrease at large momenta, but instead 
contains a constant and a slowly varying term which must be 
cancelled by appropriate choices of $\cp_q$ and $c_\NGI$. $\cp_q$ 
and $c_\NGI$ are then determined by requiring that the improved 
propagator satisfies, for a range of asymptotic values of $|p|$, 
the condition 
\begin{equation} 
\Tr \hat{S}(p) =0\, , 
\label{eq:cqfix} 
\end{equation} 
where the trace is over spin and colour indices. As $Z_q$ is just 
an overall factor in this condition, we do not need to know it in 
advance in order to determine $\cp_q$ and $c_\NGI$. The 
implementation of Eq.~(\ref{eq:cqfix}) requires that the 
asymptotic behaviour of the correlation function can be extracted 
in the region where $p^2 \ll 1/a^2$ to avoid distortions due to 
O($a^2$) errors. 
The same caveat applies to the improvement of bilinear operators discussed 
below. Exploratory numerical studies of the condition~(\ref{eq:cqfix}) 
can be found in~\cite{LUB}. 
 
Assuming that we have fixed $\cp_q$ and $c_\NGI$ in the way we have 
described, we can determine $Z_q$, by imposing, for instance, the condition 
\begin{equation} 
-i \frac1{48} \sum_\rho 
\Tr\left(\gamma_\rho {\del \over \del p_\rho} \hat{S}^{-1}(p) 
\right)\bigg|_{p^2=\mu^2} 
= 1 \,. 
\label{eq:sig1} 
\end{equation} 
In view of the above discussion it may be preferable to determine $Z_q$ using 
the forward matrix element of the conserved lattice current on quark 
states~\cite{MPSTV}. We note that, while different definitions of the 
field renormalization $Z_q$ are possible, the coefficients $\cp_q$ and 
$c_\NGI$ are completely fixed by the form of the action, and are functions 
only of $g_0^2$. This is why they are completely determined by the 
asymptotic behaviour of $\Sigma_2$. 
 
We can now see why $\cp_2$ can be set to zero in the action. The 
reason is that the term proportional to $\cp_2$, 
eq.~(\ref{eq:c2pdef}), can be removed (up to O($a^2$)) by 
performing the change of variables 
\begin{equation} 
q \to \left[1 + a \cp_2 (\rDslash+m_0)\right] q 
\,, \qquad 
\qbar \to \qbar \left[1 + a \cp_2(-\lDslash + m_0)\right] 
\,. 
\end{equation} 
The expressions for the improved quark fields in terms of the new 
variables retain the same form, Eqs.~(\ref{eq:imprquark}) 
and~(\ref{eq:impqbar}), but with a shifted improvement coefficient 
$\cp_q \to \cp_q - \cp_2$. The Jacobian for the change of 
variables shifts the $1/g_0^2$ coefficient in front of the gluon 
action, by a function of $\cp_2$ [whose expansion begins at 
O($c^{\prime 4}_2$)]. Since we are determining $\cp_q$ and 
$a(g_0)$ non-perturbatively, these shifts do not concern us. 
Similar shifts occur in the improvement coefficients, $\cp_O$, of 
other quark bilinears which we will introduce below. 
 
We have implemented the off-shell improvement procedure 
described in this sector in perturbation theory, 
using the one-loop results of Ref.~\cite{capitani}. 
Details are given in Appendix C. 
We find that a one-loop calculation can only separately 
determine the tree-level values of $\cp_q$ and $c_{NGI}$ 
(the results are $\cp_q=-1/4$ and $c_{NGI}=0$). 
At one-loop accuracy, 
only the linear combination $\cp_q+c_{NGI}$ can be determined. 
Thus, at present, the 
separate one-loop corrections to $\cp_q$ and $c_{NGI}$ are not known. 
 
\section{Improvement of bilinear quark operators} 
\label{sec:bilinears} 
 
In this section we discuss the improvement of local 
gauge-invariant bilinears. We consider only flavour non-singlet 
bilinears, but we drop flavour indices because they are not 
necessary for our discussion. We introduce the following notation: 
\begin{itemize} 
\item 
The amputated vertex function is defined as in the 
continuum~\footnote{In ref.~\cite{MPSTV} the amputated vertex function was 
denoted $\Lambda_{O_{{\Gamma}}}(p,p')$.} 
\begin{equation} 
\langle p|O_{{\Gamma}}|p'\rangle_{\amp} 
= \hat{S}(p)^{-1} G_{{\Gamma}}(p,p') \hat{S}(p')^{-1} 
\,, 
\end{equation} 
where 
\begin{equation} 
G_{{\Gamma}}(p,p') = \int d^4x d^4y e^{-ip \cdot x +ip'\cdot y} \langle 
\hat q(x) O_{{\Gamma}}(0) \hat\qbar(y) 
\rangle 
\,. 
\label{eq:gpdef} 
\end{equation} 
with $O_{{\Gamma}}(x) = \qbar(x) \Gamma q(x)$ and $\Gamma$ a Dirac matrix. 
Note that $O_{{\Gamma}}$ is defined in terms of bare quark and antiquark 
fields. We do not construct the operator using improved quark fields $\hat q$, 
because these contain the gauge non-invariant improvement term proportional 
to $c_\NGI$ which in the end, as we shall see, does not appear in bilinears. 
\item 
Improvement of bilinears requires consideration of operators of the form 
\begin{equation} 
E_\Gamma = \qbar   
\left[ 
\Gamma (\rDslash + m_0) + ( - \lDslash + m_0)\Gamma 
\right] q \,. 
\label{eq:edef} 
\end{equation} 
These operators vanish by the equations of motion 
and therefore only contribute contact terms when inserted in correlators. 
\end{itemize} 
 
It turns out that the improvement of bilinears up to O($a^2$) does not 
require the addition of gauge non-invariant operators. This is because there 
are no such operators with the same quantum numbers as the bilinears, having 
dimension 4. In fact, as mentioned above, allowed gauge non-invariant 
operators are of two types: those vanishing by the equations of motion, 
which need not be BRST invariant, and BRST invariant operators, which need 
not vanish by the equations of motion. The former appear first at dimension 
5, an example being 
\begin{equation} 
\hat\qbar 
\left[ 
\Gamma \gamma_\mu t^a A_\mu^a (\rDslash + m_0) 
+ ( - \lDslash + m_0) t^a A_\mu^a \gamma_\mu \Gamma 
\right] \hat q \,. 
\end{equation} 
The lowest dimension BRST invariant operator has, instead, dimension 6: 
\begin{equation} 
\delta_{\rm B} 
\left( \qbar[\Gamma,\gamma_\mu] t^a q \partial_\mu\bar{c}^a \right) 
= \qbar [\Gamma,\gamma_\mu] t^a q 
\left[ \partial_\mu\lambda^a + f_{abd} (\partial_\mu \bar{c}^b) c^d \right] \,. 
\end{equation} 
Our restriction to flavour non-singlet operators is crucial here. 
Improvement of flavour singlet bilinears requires, in general, 
gauge non-invariant operators of dimension 4. For example, for 
$\Gamma=1$, one requires operators of the form appearing in the 
gauge-fixing action, ${\cal{S}}_{gf}$. 
 
\subsection{Pseudoscalar density} 
 
The renormalized improved pseudoscalar density has the general form 
\begin{eqnarray} 
\widehat{P}(x) &\equiv& Z_P P(x)\,, \\ 
P(x) &=& 
\left[ \qbar(x)\gamma_5 q(x) + a \cp_P E_{\gamma_5}(x) \right] \,. 
\end{eqnarray} 
 
The form of the terms in the amputated vertex which are cancelled by an 
appropriate choice of $\cp_P$ can be found by considering the contribution 
proportional to $\cp_P$ itself. This can be obtained by calculating the 
contribution of $E_{\gamma_5}$ to the vertex function $G_P$ 
(see Eq.~(\ref{eq:gpdef})), with the result~\footnote{We stress that $G_P$ 
contains the insertion of the operator $P$ and not $\hat{P}$.} 
\begin{equation} 
a \cp_P \left[ \hat{S}(p) \gamma_5 + \gamma_5 \hat{S}(p') \right] 
+ \mbox{O}(a^2) \,. 
\end{equation} 
After amputation, one finds 
\begin{equation} 
\langle p|a\cp_P E_{\gamma_5}|p'\rangle_{\amp} = 
a\cp_P \left[ \gamma_5 \hat{S}(p')^{-1} + \hat{S}(p)^{-1} \gamma_5 \right] 
+ \mbox{O}(a^2) \,. 
\label{eq:Pcontact} 
\end{equation} 
For large $p^2$, $p'^2$ and $(p-p')^2$ we expect the continuum 
vertex function to approach its perturbative form, and thus that 
the only surviving form factor is that proportional to $\gamma_5$. 
Indeed, because of chiral symmetry, the form factor proportional 
to $\gamma_\mu\gamma_5$ vanishes in continuum perturbation theory 
when $m_q=0$. If improvement were not implemented however, a 
contribution of O($a$) proportional to 
$a(p-p')_\mu\gamma_\mu\gamma_5$ would survive, as can be seen by 
substituting $\hat{S}(p)^{-1}\sim \not \hspace{-3pt} p$ into 
Eq.~(\ref{eq:Pcontact}). This may be cancelled by tuning $\cp_P$. 
One can thus determine $\cp_P$ by setting $p\ne p'$ and imposing 
the condition 
\begin{equation} 
\lim_{|p|,|p'|,|p-p'|\to\infty} 
\Tr\left( \gamma_\mu\gamma_5 
\langle p|P|p'\rangle_{\amp} 
\right) 
= 0 \,. 
\label{eq:condP} 
\end{equation} 
An incorrect choice of $\cp_P$ would result in an O($a$) contribution 
to this trace growing like $p_\mu$. 
Eq.~(\ref{eq:condP}) is equivalent to determining $\cp_P$ from 
\begin{equation} 
- a \cp_P = \lim_{|p|,|p'|,|p-p'|\to\infty} 
{ \Tr\left(\gamma_\mu\gamma_5 
\langle p|\qbar \gamma_5 q |p'\rangle_{\amp} \right) 
\over 
\Tr\left(\gamma_\mu\gamma_5 
\langle p|E_{\gamma_5}|p'\rangle_{\amp} \right) 
} \,, 
\label{eq:cppdef} 
\end{equation} 
which shows that the overall normalization of the operator 
need not be fixed before determining $\cp_P$. 
 
The normalization constant $Z_P$ is logarithmically divergent as $a\to0$, 
and should be fixed by a renormalization condition. A possible choice is 
\begin{equation} 
Z_P \frac{1}{12} \Tr\left( \gamma_5 
\langle p|P|p\rangle_{\amp} 
\right)\bigg|_{p^2=\mu^2} = 1 \,. 
\label{eq:normP} 
\end{equation} 
We note that it is necessary to determine $\cp_P$ before $Z_P$. 
The point is that Eq.~(\ref{eq:Pcontact}) contains a term of O($a$) which is 
proportional to $\gamma_5$, coming from the ``mass term'' in the inverse 
propagator, i.e. from the part of $\hat{S}^{-1}$ proportional to the 
identity. Tuning $\cp_P$ thus changes the $\gamma_5$ form factor by terms 
of O($a$), and so changes the resulting value of $Z_P$. Note also that an 
error in the determination of $\cp_P$ leads to an error in the O($a$) part 
of $Z_P$ that is proportional to $a m_q$. Similar comments apply to other 
bilinears. 
 
\subsection{Vector Current} 
 
The improved vector current has the general form 
\begin{eqnarray} 
\hat{V}_\mu(x) &=& Z_V V_\mu(x) \,, \\ 
V_\mu(x) &=& 
\qbar(x)\gamma_\mu  q(x) 
+ a c_V \sum_\nu i \del_\nu \left[\qbar(x)\sigma_{\mu\nu}  q(x)\right] 
+ a \cp_V E_{\gamma_\mu}(x) \,. 
\label{eq:vmu} 
\end{eqnarray} 
In this case, improvement requires the determination of three constants, 
one more than for the pseudoscalar density. The new type of constant, 
$c_V$, is to be chosen to remove O($a$) contributions to physical 
amplitudes in the chiral limit. It was introduced and discussed in 
ref.~\cite{Luscher2} and determined in refs.~\cite{bcgls,guasom}. The 
contribution proportional to $\cp_V$ is needed to cancel contact terms of 
O($a$). Finally, the part of $Z_V$ proportional to $a m_q$  (usually called 
$b_V$) should be chosen so that the vector current is 
normalized as in the continuum up to O($a^2$). The normalization $Z_V$ is 
not arbitrary, unlike that of the pseudoscalar density, but should be 
fixed so that the vector current satisfies the usual current algebra 
relations. $Z_V$ is finite in the continuum limit. 
 
To determine the improvement constants $c_V$ and $\cp_V$ we use the fact 
that, for the vector current, the only continuum form factor which survives 
at large momenta is that proportional to $\gamma_\mu$. The contribution of 
$\cp_V$ to the amputated vertex,  $\langle p|V_\mu|p'\rangle_{\amp}$, is 
\begin{equation} 
a \cp_V \left[ \gamma_\mu \hat{S}(p')^{-1} 
+ \hat{S}(p)^{-1} \gamma_\mu  \right] + \mbox{O}(a^2) \,, 
\label{eq:cpvcontrib} 
\end{equation} 
which for large $|p|$, $|p'|$ and $|p-p'|$ is proportional to 
\begin{equation} 
a\cp_V \left[ \gamma_\mu \pslash' + \pslash \gamma_\mu \right] + \mbox{O}(a^2) 
= a\cp_V \left[(p+p')_\mu + i \sigma_{\mu\nu} (p-p')_\nu \right] + 
\mbox{O}(a^2) \,. 
\end{equation} 
The contribution of $c_V$, on the other hand, is asymptotically proportional 
to $\sigma_{\mu\nu} (p-p')_\nu$. In the unimproved current, therefore, 
contributions of O($a$)  proportional to the identity and to 
$\sigma_{\mu\nu}$ are not suppressed, and must be cancelled by tuning $c_V$ 
and $\cp_V$. We can determine $\cp_V$ separately by setting $p=p'$, 
since in this case the $c_V$ term does not contribute, being proportional to 
a total divergence. To isolate the unwanted form factor we project onto the 
identity matrix, and impose the asymptotic condition 
\begin{equation} 
\lim_{|p|\to\infty}\Tr\left( 
\langle p|V_\mu|p\rangle_{\amp} 
\right) = 0 \,. 
\label{eq:detcvp} 
\end{equation} 
Having determined $\cp_V$, we then fix $c_V$ by requiring that 
the tensor form factor is absent asymptotically, i.e. 
\begin{equation} 
\lim_{|p|,|p'|,|p-p'|\to\infty} 
\Tr\left( \sigma_{\mu\nu}\langle p|V_\mu|p'\rangle_{\amp} \right) = 0 \,. 
\label{eq:detcv} 
\end{equation} 
Here and in the following, there is no implied summation over 
repeated indices unless explicitly indicated. 
Since the tuning of $\cp_V$ affects both scalar and tensor form factors, 
it is necessary that the conditions~(\ref{eq:detcvp}) and~(\ref{eq:detcv}) be 
applied in the specified order. 
 
Finally, the normalization constant $Z_V$ can be obtained 
by imposing the WTI 
\begin{equation} 
Z_V  \sum_\mu \langle p|\tilde\del_\mu V_\mu |p'\rangle_{\amp} 
= \hat{S}(p)^{-1} - \hat{S}(p')^{-1} \,, 
\label{eq:fixZVbyWI} 
\end{equation} 
where $\tilde\del_\mu$ is the symmetric (improved) lattice derivative 
\begin{equation} 
\tilde\del_\mu f(x) = \frac1{2a} \left[f(x\!+\!a\hat\mu) 
- f(x\!-\! a\hat\mu) \right]\,. 
\end{equation} 
As discussed in ref.~\cite{MPSTV}, this gives the same result as the condition 
\begin{equation} 
Z_V  \frac1{12} \Tr\left( \gamma_\nu 
\langle p|V_\nu|p\rangle_{\amp} 
\right)\bigg|_{p^2=\mu^2} = 1 \,. 
\label{eq:normV2} 
\end{equation} 
Note that, if we express all improved quark fields in terms of 
bare quark fields, using Eqs.~(\ref{eq:imprquark}) and~(\ref{eq:impqbar}), 
both sides of Eq.~(\ref{eq:fixZVbyWI}) will be proportional to $Z_q$. 
Thus we do not to need to know $Z_q$ in order to determine 
$Z_V$ using the WTI. 
The factors of $Z_q$ simply ensure that each term in 
Eq.~(\ref{eq:fixZVbyWI}) remains finite in the continuum limit. 
 
An alternative way of determining $Z_V$ is to require 
that the charge of a particular hadron equals its physical value. 
In practice, this may be preferable to the methods described above, 
which are based on quark correlators, 
since determinations using hadronic states typically have smaller 
errors. 
 
\subsection{Axial Current} 
 
Improvement of the axial current is accomplished following similar 
steps to those needed for the vector current. 
The improved form of the axial current is 
\begin{eqnarray} 
\hat{A}_\mu(x) &=& Z_A  A_\mu(x) \,, \\ 
A_\mu(x) &=& 
\qbar(x)\gamma_\mu \gamma_5  q(x) 
+ a c_A \del_\mu \left[\qbar(x) \gamma_5  q(x)\right] 
+ a \cp_A E_{\gamma_\mu\gamma_5}(x) \,. 
\label{eq:amudef} 
\end{eqnarray} 
In this case, the contribution of $E_{\gamma_\mu\gamma_5}$ 
to the amputated vertex, 
$\langle p|A_\mu |p'\rangle_{\amp}$, is 
\begin{equation} 
a \cp_A \left[ \gamma_\mu\gamma_5 \hat{S}(p')^{-1} + 
\hat{S}(p)^{-1} \gamma_\mu\gamma_5 \right] + \mbox{O}(a^2) \,. 
\end{equation} 
The only form factor which must survive asymptotically in the continuum 
is that proportional to $\gamma_\mu\gamma_5$. 
We therefore first determine $\cp_A$ from the condition 
\begin{equation} 
\lim_{|p|\to\infty}\Tr\left(\sigma_{\mu\nu}\gamma_5 
\langle p|A_\mu|p\rangle_{\amp} 
\right) = 0 
\end{equation} 
and then fix $c_A$ by imposing 
\begin{equation} 
\lim_{|p|,|p'|,|p-p'|\to\infty} 
\Tr\left( \gamma_5 
\langle p|A_\mu|p'\rangle_{\amp} 
\right) = 0 \,. 
\end{equation} 
Finally, the normalization constant $Z_A$ can be determined 
by enforcing the WTI 
\begin{equation} 
Z_A  \langle p|\sum_\mu\tilde\del_\mu A_\mu|p'\rangle_{\amp} 
- 2 \hat{m}_q Z_P \langle p|P|p'\rangle_{\amp} 
= - \left[ \hat{S}(p')^{-1}\gamma_5 + \gamma_5 \hat{S}(p)^{-1} \right] \,. 
\label{eq:axialWI} 
\end{equation} 
For $p=p'$ this identity fixes the value of $\hat{m}_q$, 
since the first term on the l.h.s. vanishes. 
Here, $\hat{m}_q$ is a possible definition of the renormalized quark mass, 
which is automatically improved to O($a^2$), 
being determined from quantities that are already improved at this order. 
Note that the combination $\hat{m}_q Z_P$ is determined unambiguously, 
but that, as in the continuum, the value of $\hat{m}_q$ depends on 
the normalization condition chosen for $Z_P$. 
Having found $\hat{m}_q$, 
$Z_A$ can then be obtained by considering $p\ne p'$, and tracing 
Eq.~(\ref{eq:axialWI}) with $\gamma_\mu\gamma_5$. 
As for $Z_V$, the determinations of $\hat{m}_q$ and 
$Z_A$ do not require knowledge of $Z_q$. 
 
\subsection{Scalar and tensor bilinears} 
 
Finally, we briefly sketch the analysis for the scalar and tensor operators. 
The improved scalar density is 
\begin{eqnarray} 
\hat{S}(x) &=& Z_S  S(x) \,,\\ 
S(x) &=& \qbar(x)  q(x) 
+ a \cp_S E_{S}(x) \,. 
\end{eqnarray} 
We can fix $\cp_S$ from the condition 
\begin{equation} 
\lim_{|p|\to\infty}\Tr\left(\gamma_\mu 
\langle p|S|p\rangle_{\amp} \right) = 0 
\end{equation} 
and then determine $Z_S$ by imposing 
\begin{equation} 
Z_S  \frac1{12}\Tr\left( 
\langle p|S|p\rangle_{\amp} 
\right)\bigg|_{p^2=\mu^2} = 1 
\,. 
\end{equation} 
 
The improved tensor operator has the form 
\begin{eqnarray} 
\hat{T}_{\mu\nu}(x) &=& Z_T  T_{\mu\nu}(x) \,, \\ 
T_{\mu\nu}(x) &=& i\qbar(x)\sigma_{\mu\nu}  q(x) 
+ a c_T (\del_\mu V_\nu - \del_\nu V_\mu) 
+ a \cp_T E_{T_{\mu\nu}}(x) \,, 
\end{eqnarray} 
where in $V_\mu$, defined in Eq.~(\ref{eq:vmu}), one does not need to 
include the $O(a)$ terms proportional to $c_V$ and $\cp_V$. We can obtain 
$\cp_T$ by requiring 
\begin{equation} 
\lim_{|p|\to\infty}\Tr\left(\gamma_\rho\gamma_5 
\langle p|T_{\mu\nu}|p\rangle_{\amp} \right) = 0 
\,, 
\end{equation} 
where all Lorentz indices are different, and then determine $c_T$ by 
enforcing the condition 
\begin{equation} 
\lim_{|p|,|p'|,|p-p'|\to\infty} 
\Tr\left( \gamma_\mu 
\langle p|T_{\mu\nu}|p'\rangle_{\amp} 
\right) = 0 \,. 
\end{equation} 
Finally, we can fix $Z_T$ by imposing 
\begin{equation} 
Z_T \frac1{12}\Tr\left( \sigma_{\nu\rho} 
\langle p|T_{\nu\rho}|p\rangle_{\amp} 
\right)\bigg|_{p^2=\mu^2} = 1 \,. 
\end{equation} 
 
\section{Conclusions} 
\label{sec:conclusions} 
 
We have presented a method for determining the off-shell 
improvement coefficients of gauge invariant bilinears in the 
massive case, using quark correlation functions. We stress that 
our ultimate interest is in the calculation of the coefficients 
necessary for the improvement of on-shell quantities, but that we 
might wish to use off-shell correlators as an intermediate step. A 
detailed numerical study will be necessary to determine whether 
the method is practical. The pilot study of ref.~\cite{LUB} 
suggests that it will be difficult to separate $\cp_q$ from 
$c_{NGI}$, in which case the method works only in the chiral 
limit. 
 
The whole approach can be extended to the case of non-degenerate 
quark masses, which is relevant for heavy flavour  phenomenology. 
The only change is in the form of the WTI's. For the axial WTI, 
$2\hat{m}_q$ should be replaced by $\hat{m}_1+\hat{m}_2$, while 
for the vector identity, a term proportional to the mass 
difference should be added. 
 
The approach can also be extended to deal with more complicated composite 
operators, such as four-quark operators relevant to study the effective weak 
Hamiltonian, though the number of operators which must be included may 
render the procedure impractical. 
 
We stress again, however, that our approach fails, in general, for 
operators which can mix with lower dimensional ones (such as the 
octet part of the weak Hamiltonian). In this situation, lower 
dimensional operators may give rise in correlators to a large 
momentum behaviour that cannot be distinguished, and thus 
disentangled, from that coming from the exchange of Goldstone 
bosons in correlators with the insertion of the original operator. 
 
\section*{Appendix A} 
 
In this Appendix we want to explain in a simple example why, in presence of 
spontaneous breaking of chiral symmetry, it may not be possible to fix 
non-perturbatively all the coefficients of the mixing of lower dimensional 
operators with higher dimensional ones by simply looking at the high momentum 
behaviour of correlators. The reason is that correlators with the insertion of 
lower dimensional operators may turn out to have the same high momentum 
behaviour as that coming from contributions due to the exchange of Goldstone 
bosons to (some of the) correlators in which the original operator is 
inserted. 
 
To prove this statement let us consider the correlator~\footnote{As in the 
rest of the paper also in this Appendix to simplify notations we will 
neglect flavour indices.} 
\begin{equation} 
G^{A}_{d}(p,p') \equiv \int d^{4}x d^{4}y e^{-ipx+ip'y} \langle 
\delta_{A}\Big{(}\bar q (y) q(x) O_{d}(0)\Big{)}  \rangle\, , 
\label{eq:uno} 
\end{equation} 
where $O_d$ is an operator of dimension $d$ and $\delta_{A}(O)$ represents 
the axial variation of the operator $O$. A simple dimensional counting shows 
that the asymptotic scaling behaviour of $G^{A}_{d}(p,p')$ for large $p$ and 
$p'$ is given by the formula 
\begin{equation} G_{d}(\lambda p,\lambda p') 
\mathop \sim \limits_{\lambda \to \infty } \lambda ^{d-5}\, . 
\label{eq:due} \end{equation} 
A similar analysis for the Green function 
\begin{equation} 
G^{m}_{d}(p,p') \equiv 
m \int d^{4}x \, d^{4}y \, d^{4}z \, e^{-ipx+ip'y} \langle 
\bar q(z)\gamma_5 q(z) \bar q (y) q(x) O_{d}(0) \rangle\, , 
\label{eq:tre} 
\end{equation} 
in which the operator $P=\bar q\gamma_5 q$ is inserted together with $O_d$, 
gives 
\begin{equation} 
G^{m}_{d}(\lambda p,\lambda p') \mathop \sim \limits_{\lambda \to \infty } 
\lambda ^{d-6}\, . 
\label{eq:quattro} 
\end{equation} 
 
Suppose now we are interested in the case of the mixing of an operator of 
dimension 6 (like the octet part of the effective weak Hamiltonian) with 
an operator of dimension 3 (like $\bar q\gamma_5 q$ or $\bar q q$). 
	From the scaling behaviours~(\ref{eq:due}) and~(\ref{eq:quattro}) we 
immediately conclude that, in the absence of a massless pion, everything 
works fine, because at large momenta 
 
1) the contribution of the correlator $G^{m}_{d}$ to the relevant axial WTI 
is subleading with respect to that coming from $G^{A}_{d}$ both for $d=6$ 
and $d=3$; 
 
2) correlators with inserted operators of dimension 6 and 3 have well 
separated asymptotic behaviours. 
 
In this circumstance the unwanted contributions (i.e. those corresponding to 
$d=3$ in the above scaling laws) can be isolated in the various correlators 
and set to zero by appropriately choosing the mixing coefficients at our 
disposal. 
 
The problem arises if chirality is spontaneously broken, because in the 
chiral limit the pion contributes to $G^{m}_{6}$ a term 
\begin{eqnarray} 
&\!&G^{m}_{6}(p,p') \Big {|}_{pion} = f_{\pi} 
\int d^{4}x \, d^{4}y \, e^{-ipx+ip'y} \langle \pi|\bar q (y) 
q(x) O_{6}(0) \rangle\nonumber \\ 
&\!&\mathop \sim \limits_{p,p' \to \infty} 
\int d^{4}x \, d^{4}y \, e^{-ipx+ip'y}c_{W}(x,y) \langle \pi|\bar q 
\gamma_{5} q(0) \rangle\, , 
\label{eq:cinque} 
\end{eqnarray} 
where $f_\pi$ is pion decay constant. The first equality in 
Eq.~(\ref{eq:cinque}) follows from PCAC, the second from inserting the 
leading non-vanishing contribution of the O.P.E. 
\begin{equation} 
\bar q (y)  q(x) O_{6}(0)\mathop \sim \limits_{x\sim y \sim 0} 
c_{W}(x,y) \bar q 
\gamma_{5} q(0) + \ldots \, .\label{sei} 
\end{equation} 
A straightforward dimensional argument leads to the asymptotic scaling law 
\begin{equation} 
c_{W}(\lambda p,\lambda p') = \int d^{4}x \, d^{4}y \, e^{-i\lambda 
px+i\lambda p'y}c_{W}(x,y) \mathop \sim \limits_{\lambda \to \infty } 
\lambda ^{-2}\, . 
\label{eq:sette} 
\end{equation} 
The trouble with this behaviour is that it is the same as that predicted by 
Eq.~(\ref{eq:due}) for an operator of dimension $d=3$. This simple 
observation proves our theorem. In fact, in the ``kinematical'' situation of 
the example we are considering, it would be impossible to distinguish at high 
momenta the contribution to the axial WTI coming from an operator of 
dimension 3 from the similar one due to the exchange of a massless pion in 
$G^{m}_{6}$. 
 
\section*{Appendix B} 
 
In this appendix we wish to give a simple argument to show that in a 
gauge-fixed theory BRST symmetry does not forbid the mixing of gauge 
invariant operators with gauge non-invariant ones if the latter vanish by 
the equations of motion~\cite{zuber,BBH}. 
 
To simplify notations we collectively indicate by $[\phi_k(x)]$ the set of 
fundamental (fermionic and bosonic) fields, including ghosts, appearing 
in the Lagrangian of the theory and by $[B_k(x)]$ their (bosonic and 
fermionic) BRST variations~\footnote{Throughout this appendix to avoid 
cumbersome equations we will be rather cavalier about signs.}: 
\begin{equation} 
\delta_{\rm B} \phi_k(x)= B_k(x)\, . 
\label{eq:BRSTPHI} 
\end{equation} 
 
Introducing the sources $J^k_B(x)$, coupled to the variations $B_k(x)$, we 
define the so-called Zinn-Justin functional by 
\begin{eqnarray} 
&\,&{\cal{Z}}[j,J_B,\eta]\equiv \int {\cal{D} \phi}\,\exp \{-{\cal{S}}[\phi]+ 
\int {\phi_k j^k }+\int {J^k_BB_k}+\int {\eta O}\}\equiv \nonumber \\ 
\nonumber\\ 
&\,&\equiv \int {\cal{D} \phi}\,\exp \{{\cal{A}}[\phi ,J_B]+\int {\phi_k 
j^k}+ \int {\eta O}\} \, , 
\label{eq:ZJF} 
\end{eqnarray} 
where $\eta$ is the source coupled to the BRST-invariant operator $O$ and a 
summation over repeated indices is understood. 
Expectation values with respect to this measure will be denoted 
\begin{equation} 
\left\langle \right\rangle\Big{|}_{j,J_B,\eta}\, 
\,. 
\end{equation} 
In the second equality of 
Eq.~(\ref{eq:ZJF}) we have introduced the definition 
\begin{equation} 
{\cal{A}}[\phi,J_B]\equiv - {\cal{S}}[\phi]+\int {J^k_B B_k} \,. 
\label{eq:ZJA}\end{equation} 
We assume that the operator $O$ is renormalized, 
so that its insertions with the fundamental fields of the theory 
are finite if the fields and $O$ are physically separated. 
The issue at hand is whether there can be contact terms, 
whose coefficients diverge when the regularization is removed, 
and the removal of which requires the use of gauge non-invariant operators. 
 
All the consequences of the BRST symmetry can be compactly expressed 
by the so-called Zinn-Justin equation 
\begin{equation} 
\int dx {{{\delta {\cal{Z}}[j,J_B,\eta]} \over {\delta J^k_B(x)}}j^k(x)}=0\, . 
\label{eq:ZJEQ} 
\end{equation} 
This is derived by performing the change of variables 
induced by the BRST transformations~(\ref{eq:BRSTPHI}) 
in the functional integral~(\ref{eq:ZJF}), 
assuming that the Jacobian is unity. 
It is useful to expose the condition on $B_k$ that a unit Jacobian implies: 
\begin{eqnarray} 
0&=& 
\int dx \int{\cal{D} \phi }\,{\delta \over \delta \phi_k(x)} 
\left\{ B_k(x) 
\exp \{{\cal{A}}[\phi ,J_B]+\int {\phi_k j^k}+ \int 
{\eta O}\} \right\} \nonumber \\ 
&=& 
\left\langle \int dx {\delta B_k(x)\over \delta \phi_k(x)}\right\rangle 
\Big{|}_{j,J_B,\eta}\, 
+ 
\left\langle \int dx B_k(x) j_k(x)\right\rangle 
\Big{|}_{j,J_B,\eta}\, 
\label{eq:ZJEQQ} 
\end{eqnarray} 
where we have used the BRST invariance of the action, 
\begin{equation} 
0=\delta_{\rm B} {\cal{S}}=\int dx B_k(x) \frac{\delta {\cal{S}}}{\delta 
\phi_k(x)} \,, \label{eq:SW} 
\end{equation} 
and the nilpotency of BRST transformations, 
\begin{equation} 
0= \int dy  B_j(y) \frac{\delta B_k (x)}{\delta \phi_j(y)}\, . 
\label{eq:BW} 
\end{equation} 
The Zinn-Justin equation~(\ref{eq:ZJEQ}) follows from~(\ref{eq:ZJEQQ}) 
as long as the condition for unit Jacobian, 
\begin{equation} 
\left\langle\int\,dx\frac{\delta B_k(x)} {\delta \phi_k(x)} 
\right\rangle \Big{|}_{j,J_B,\eta}=0 
\,, \label{eq:BRSTID} 
\end{equation} 
is satisfied. 
One readily verifies that antisymmetry arguments imply that 
this equation is trivially satisfied. 
Indeed, the BRST variations, $B_k$, are such that 
\begin{equation} 
\frac{\delta B_k(x)} {\delta \phi_k(x)}\equiv 0\, . 
\label{eq:ID} 
\end{equation} 
 
We note that the Zinn-Justin equation holds 
in a regularized theory as long as regularized BRST transformations 
can be  defined under which the regularized action is invariant, 
and which satisfy~(\ref{eq:BRSTID}). 
This is the case for lattice QCD~\cite{LuscherBRST}(see sec.~2.1). 
 
The regularized Green functions of the operator $O$ obey the 
Slavnov-Taylor identities that follow from Eq.~(\ref{eq:ZJEQ}). 
If these Green functions contain contact terms which diverge when 
the regulator is removed, then these contact terms must separately 
satisfy the identities. Thus we can choose the counter-terms we add 
to cancel these divergences so that the identities, and thus the 
Zinn-Justin equation itself, are still satisfied. 
Thus we introduce the modified functional 
\begin{equation} 
\tilde {{\cal{Z}}}[j,J_B,\eta]\equiv \int {\cal{D} \phi}\,\exp 
\{{\cal{A}}[\phi ,J_B] + \int {\phi_k j^k}+\int {\eta (O+F[J_B, \phi]})\}  \, 
,   \label{eq:ZJIDM} 
\end{equation} 
where $F[J_B, \phi]$ is a local functional of the sources, $J^k_B$, and the 
fundamental fields, $\phi_k$, and derive the constraints imposed on 
the form of the functional $F$ by the modified Zinn-Justin equation 
\begin{equation} 
\int dx {{{\delta \tilde{{\cal{Z}}}[j,J_B,\eta]} 
\over {\delta J^k_B(x)}}j^k(x)}=0\, . 
\label{eq:ZJTEQ} 
\end{equation} 
The significance of the dependence of $F$ on $J^k_B$ will become apparent 
in the concrete example we give below. 
 
A little calculation shows that~(\ref{eq:ZJTEQ}) is equivalent to the 
equation 
\begin{eqnarray} 
&\,&\Big{\langle} \int dy \,\eta (y) \Big{[}\int dx \frac{\delta^2 F(y)} 
{\delta J^k_B(x) \delta \phi_k(x)} +  WF(y) \Big{]} +  \label{eq:ZTJEQ}\\ 
&\,& + \int dy \,\eta (y) \int  dy' \eta (y') \int dx 
\frac{\delta F(y)}{\delta J^k_B (x)} 
\frac{\delta(O(y')+F(y'))}{\delta \phi_k (x)}\Big{\rangle} 
\Big{|}_{j,J_B,\eta}  =0 \, ,\nonumber 
\end{eqnarray} 
where we have introduced the operator 
\begin{eqnarray} 
&\,&W = \int  dx \Big{(} \frac{\delta {\cal{A}}} {\delta \phi_k (x)} 
\frac{\delta} {\delta J^k_B(x)} + \frac{\delta {\cal{A}}} {\delta J^k_B(x)} 
\frac{\delta}{\delta \phi_k (x)} \Big{)} = \nonumber \\ 
&\,& =\int  dx\Big{(}\frac{\delta {\cal{A}}}{\delta \phi_k (x)} 
\frac{\delta} {\delta J^k_B(x)}+ B_k(x)\frac{\delta} {\delta \phi_k (x)} 
\Big{)} \, , 
\label{eq:WDEF} 
\end{eqnarray} 
so that 
\begin{equation} 
W F(x) = \int dy{{\delta {\cal{A}}} \over {\delta \phi_k (y)}} \frac{\delta 
F(x)} {\delta J^k_B(y)}  +\delta_{\rm B} F(x)\, . 
\label{eq:W} 
\end{equation} 
To get Eq.~(\ref{eq:ZTJEQ}) we have exploited the identity~(\ref{eq:ID}) and 
the BRST-invariance of the generalized action (Eq.~(\ref{eq:ZJA})), which 
in our notation amounts to the two equations~(\ref{eq:SW}) and~(\ref{eq:BW}). 
Limiting for simplicity the analysis to Green functions with single insertions 
of the operator $O$, Eq.~(\ref{eq:ZTJEQ}) implies the condition 
\begin{equation} 
WF(x)+\hbar\int dy \frac{\delta^2 F(x)}{\delta \phi_k(y)\delta J^k_B(y)} 
\equiv W' F(x) = 0\, , 
\label{eq:WAUX} 
\end{equation} 
where we have reinstated the dependence on $\hbar$, and defined the new 
operator $W'$. 
 
The key observation needed to solve~(\ref{eq:WAUX}) 
is to recognize~\cite{zuber} that $W$ is a nilpotent operator, 
$W^2=0$. Using this property, and the fact the 
fermionic operator  $\int dy {\delta^2 }/{\delta J^k_B(y) \delta \phi_k(y)}$ 
is nilpotent, it then follows that $W'$ itself is nilpotent, 
$W'^2=0$. Thus a solution to~(\ref{eq:WAUX}) is $F = W' C$. 
To show that this is the only solution, to all orders in 
perturbation theory, we expand $F$ in powers of $\hbar$: 
$F=F_0+\hbar F_1+\hbar^2F_2+\hbar^3F_3+\ldots$. 
Introducing this expansion in Eq.~(\ref{eq:WAUX}) and repeatedly 
using the nilpotency of $W$ and $W'$, one obtains the 
solution in the form\footnote{%
To begin the iteration one needs the result that 
solutions to $W G = 0$ are of the form $G= W H$, 
where $H$ is a local functional of fields and 
sources~\cite{zuber}.} 
\begin{eqnarray} 
&\,&F=W(C_0+\hbar C_1+\hbar^2C_2+\hbar^3C_3+\ldots)+ \nonumber \\ 
&\,&+\hbar\int dy\frac{\delta^2}{\delta \phi_k(y) \delta 
J_B^k(y)}(C_0+\hbar C_1+\hbar^2C_2+\ldots)\, , 
\label{eq:SOL} 
\end{eqnarray} 
where $C_i[\phi,J_B]$ are arbitrary local functionals of the sources, 
$J_B$, and the fundamental fields of the theory. 
The series expansion 
in $\hbar$ in the r.h.s. of Eq.~(\ref{eq:SOL}) can be formally resummed, 
giving 
\begin{equation} 
F(x)= W' C(x) = W C(x)+\hbar\int dy\frac{\delta^2 C(x)}{\delta \phi_k(y) 
\delta J_B^k(y)}\, . 
\label{eq:SOLFIN} 
\end{equation} 
 
Remembering the form of $W$ (see Eq.~(\ref{eq:WDEF})), we can split $F$ in a 
BRST-invariant and a non BRST-invariant part: 
\begin{equation} 
F(x)=F_{NB}(x)+\delta_{\rm B} C(x) 
\label{eq:SPLIT} 
\end{equation} 
\begin{equation} 
F_{NB}(x)=\int dy{{\delta A} \over {\delta \phi_k (y)}} \frac{\delta C(x)} 
{\delta J^k_B(y)}+  \hbar\int dy\frac{\delta^2 C(x)} {\delta 
\phi_k(y)\delta J_B^k(y)}\, . 
\label{eq:FNB} 
\end{equation} 
We are really interested in $F_{NB}$, 
because the BRST-invariant part of $F$ can always be 
reabsorbed in a redefinition of $O$. 
We can now see that, as expected, the insertion of $F_{NB}$ 
with the fundamental fields of the theory gives rise only to contact 
terms, thanks to the locality of the functional $C$. To see this we 
consider the Green function 
\begin{eqnarray} 
&\,& G(x,x_1,\ldots , x_n) = 
\label{eq:G2} \\ &\,& =\int 
{\cal{D} \phi }\,{\rm e}^{{\cal{A}}[\phi,J_B]/{\hbar}} 
F_{NB}(x) \phi_{k_1}(x_1)\ldots\phi_{k_n}(x_n) \, . \nonumber 
\end{eqnarray} 
Inserting Eq.~(\ref{eq:FNB}) one gets, 
after an integration by parts, 
\begin{eqnarray} 
&\,& G(x,x_1,\ldots , x_n) = \label{eq:G1} \\ &\,& = -\hbar 
\sum_{i=1}^n  \delta (x-x_i) \left\langle \phi_{k_1}(x_1)\ldots X_{k_i}(x_i) 
\ldots\phi_{k_n}(x_n) \right\rangle\, , 
\nonumber 
\end{eqnarray} 
where we have used the definition 
\begin{equation} 
\frac{\delta C(x)}{\delta J_B^{k_i}(x_i)}\equiv X_{k_i}(x)\delta(x-x_i)\, . 
\label{eq:XDEF} 
\end{equation} 
If we now set $J_B=0$ in Eq.~(\ref{eq:G1}) we see explicitly 
that $F_{NB}$ generates contact terms, and that they are not constrained by 
BRST symmetry (since $C$ and thus the $X_k$ are arbitrary). 
 
Returning to the full expression of $F$, 
we can now rewrite Eq.~(\ref{eq:SOLFIN}) as 
\begin{equation} 
F(x) = \frac{\delta {\cal{A}}} {\delta \phi_k (x)}X_k(x) 
+\delta_{\rm B} C(x) + \hbar \frac{\delta X_k(x)}{\delta \phi_k (x)}\, . 
\label{eq:WC} 
\end{equation} 
Apart from the last term, Eq.~(\ref{eq:WC}) proves our initial statement, 
i.e.~$F$ vanishes by the equations of motion up to BRST-invariant terms. 
The last term is proportional to $\delta(0)$. It is present to ensure 
that the term vanishing by the equations of motion does in fact lead to 
a contact term [as seen in the derivation of Eq.~(\ref{eq:G1}) above]. 
More formally, 
it compensates divergent contact terms (hidden) in the 
product of operators appearing in the first term of Eq.~(\ref{eq:WC}). 
For the flavor non-singlet operators considered in this work, 
one can in fact show that this term is algebraically zero. 
 
We want to end this appendix by clarifying, in a significant example, the 
physical meaning of the presence of the counter-term $F$ in 
Eq.~(\ref{eq:ZJIDM}). For this purpose let us consider the 
Slavnov-Taylor identity 
\begin{equation} 
\left.\Big{[}{{{\delta ^3\tilde {\cal{Z}}} \over \delta j^n(z) 
{\delta J^m_B(t) 
\delta\eta (x)}}} +{{{\delta ^3\tilde {\cal{Z}}} \over \delta j^m(t) {\delta 
J^n_B(z) \delta\eta (x)}}}\Big{]}\right |_{j=J_B=\eta =0} =0\, , 
\label{eq:ZJTILDE} \end{equation} 
that follows from equation~(\ref{eq:ZJTEQ}) by successively 
differentiating with respect to $j_m(t)$, $j_n(z)$ and $\eta(x)$. 
After some algebra one gets~\footnote{To avoid introducing a new symbol we 
now use $X_k$ to indicate the functional derivative $\delta C/\delta 
J^k_B$ {\em evaluated at $J_B=0$}.} 
\begin{eqnarray} 
&\,&\left.{{{\delta 
^3\tilde {\cal{Z}}} \over \delta j^n(z) {\delta J^m_B(t) \delta\eta 
(x)}}}\right|_{j=J_B=\eta =0} =\left\langle {\phi_n(z)B_m(t)O(x)} 
\right\rangle + \label{eq:OBPHI1} \\ &\,& + \delta (x-z)\left\langle 
X_n(z)B_m(t)\right\rangle+ \delta (x-t)\left\langle 
\phi_n(z){\delta_{\rm B} (X_m(t))}\right\rangle+ \nonumber \\ &\,& + 
\delta (x-z)\delta (x-t)\left\langle X_{nm}(x)\right\rangle\nonumber 
\end{eqnarray} 
\begin{eqnarray} 
&\,&\left.{{{\delta ^3\tilde {\cal{Z}}} \over \delta j^m(t) {\delta 
J^n_B(z) \delta\eta (x)}}}\right|_{j=J_B=\eta =0}= 
\left\langle {\phi_m(t)B_n(z)O(x)} \right\rangle + 
\label{eq:OBPHI2}\\ &\,&+ 
\delta (x-t)\left\langle X_m(t)B_n(z)\right\rangle+ 
\delta (x-z)\left\langle \phi_m(t){\delta_{\rm B}(X_n(z))}\right\rangle+ 
\nonumber \\ &\,& + 
\delta (x-z)\delta (x-t)\left\langle X_{mn}(x)\right\rangle\, .\nonumber 
\end{eqnarray} 
In the above equations we have used the formal expansion~\footnote{The term 
with $n=0$ is dropped from the expansion because a $J_B$-independent piece 
would give a BRST-invariant contribution to $O$.} 
\begin{equation} 
C(x)=\sum^\infty_{n=1} J^{k_1}_B(x)\ldots J^{k_1}_B(x) X_{k_1\ldots 
k_n}(x)\, . 
\label{eq:CDEF} 
\end{equation} 
In Eqs.~(\ref{eq:OBPHI1}) and~(\ref{eq:OBPHI2}) we see explicitly 
the meaning of the terms in $F_{NB}$ proportional to $J_B$ 
(which arise in part from terms in $C$ quadratic in $J_B$): 
they lead to the double contact terms proportional to $X_{nm}$. 
Notice that the sum of all contact terms adds up to zero by itself 
by virtue of the (anti)symmetry properties of $X_{nm}$ and of the BRST 
identity obeyed by $\left\langle {B_n(z)X_m(t)} \right\rangle $, which 
reads 
\begin{equation} 
\left\langle {X_m(t)B_n(z)} \right\rangle= 
\left\langle {\delta_{\rm B} (\phi_n (z))X_m(t)}\right\rangle= 
-\left\langle {\phi_n (z)\delta_{\rm B} (X_m(t))}\right\rangle \, . 
\label{eq:OBBRST} 
\end{equation} 
The cancellation of contact terms is also consistent with the equation 
\begin{equation} 
\left\langle {B_m(t)\phi_n (z)O(x)} \right\rangle + 
\left\langle {\phi_m (t)B_n(z)O(x)} \right\rangle =\left\langle 
{\delta_{\rm B} (\phi_m(t)\phi_n (z)O(x))}\right\rangle=0 
\label{eq:OIDD} 
\end{equation} 
that follows from the BRST invariance of $O$. 
 
Let us summarize what we have learned. It is consistent with BRST 
invariance for the gauge-invariant operator to have (divergent) 
contact terms with the fundamental fields. These can be cancelled 
by appropriate choices of the $X_n$ and $X_{nm}$ in such a way that the 
resulting correlation functions have finite Fourier transforms. 
In other words, adding the counter-term $F$ amounts to redefining 
the correlation functions as follows: 
\begin{eqnarray} 
&\,&\left\langle {\phi_m (t)B_n(z)O(x)} \right\rangle ' \equiv 
\left\langle {\phi_m (t)B_n(z)O(x)}\right\rangle +\nonumber\\ 
 &\,& +\delta (x-z)\left\langle {\phi_m(t)\delta_{\rm B} (X_n(z))} 
\right\rangle +\delta (x-t)\left\langle X_m(t)B_n(t)\right\rangle 
+\nonumber\\ &\,& + 
\delta (x-z)\delta (x-t)\left\langle X_{mn}(x)\right\rangle\, , 
\label{eq:OBPHIP} 
\end{eqnarray} 
so that they are finite for all $x,z$ and $t$. 
The Slavnov-Taylor identity~(\ref{eq:ZJTILDE}) then implies that 
\begin{equation} 
\left\langle {\phi_m (t)B_n(z)O(x)} \right\rangle ' 
+\left\langle {\phi_n (z)B_m(t)O(x)} \right\rangle '= 0\, , 
\label{eq:OBPHIPP} 
\end{equation} 
i.e.~$\left\langle {\phi_m (t)B_n(z)O(x)} \right\rangle'$ and 
$\left\langle {\phi_m (t)B_n(z)O(x)} \right\rangle$ satisfy the same BRST 
identity. 
 
What are the constraints on the counter-term implied by BRST invariance? 
Although $X_n$, 
the contact term between $\phi_n$ and $O$, is not constrained, 
the contact term between $\delta_B \phi_n$ and $O$ is required 
to be $\delta_B X_n$. 
This relationship between contact terms is encoded in the $J_B$ 
dependence of $F$. 
If more than two fields collide, then additional arbitrary contact 
terms are allowed. 
This way of understanding the result 
makes it clear that the general form of the 
contact terms could, with a lot of imagination, have been guessed 
without reference to the Zinn-Justin equation. 
 
\section*{Appendix C} 
\newcommand{\Aslash}{\not \hspace{-5pt} A} 
 
In this appendix we test our method for off-shell improvement using 
one-loop perturbation theory. We focus on the quark propagator, 
the improvement of which requires, we claim, 
the gauge non-invariant term proportional to $c_{NGI}$. 
The necessary one-loop perturbative calculations have been worked out 
by Capitani {\em et al}, and presented in final form in Ref.~\cite{capitani}. 
We stress, however, that the method of off-shell improvement 
presented in Ref.~\cite{capitani} does not include the $c_{NGI}$ term. It
works, however, up to one-loop, because, as we shall show in 
this Appendix, $c_{NGI}$ actually vanishes at tree-level.
 
It is convenient to work with the inverse bare lattice propagator since this 
is the quantity calculated in perturbation theory. 
Thus we rewrite Eq.~(\ref{eq:impS}) as 
\begin{equation} 
{\hat{S}^{-1}(p) \over Z_q^0} = S_L^{-1}(p) 
\left(1 + a b_q m - 2 a \cp_q S_L^{-1}(p) 
-2 a c_\NGI i\pslash + \mbox{O}(a^2) \right) \,. 
\label{eq:impSinv} 
\end{equation} 
Here we have used the fact that $[S_L^{-1},\pslash]=0$ 
(since the only gamma-matrix structure allowed in $S_L^{-1}$ is $\pslash$), 
and expanded the quark-field normalization factor in the conventional way 
\begin{equation} 
Z_q = Z_q^0 (1 + a b_q m) \,. 
\end{equation} 
The advantage of moving $Z_q^0$ to the l.h.s. of 
Eq.~(\ref{eq:impSinv}) is that it separates the issues of improvement 
and normalization. The l.h.s. is proportional, up to $O(a^2)$, to 
the continuum inverse propagator, and thus it should be possible 
to choose $b_q$, $\cp_q$ and $c_{NGI}$ so that the r.h.s. 
does not contain $O(a)$ terms. 
An important constraint is provided by the fact that 
these three improvement coefficients should be functions only of 
the bare coupling $g_0^2$.\footnote{%
Strictly speaking, improvement coefficients are functions of 
the coupling $g_0^2(1+b_g am)$, as explained in Ref.~\cite{Luscher2}. 
This distinction is, however, important only at two-loop order for 
the quark propagator.} 
 
The one-loop result for the inverse propagator with an improved action 
takes the form 
\begin{eqnarray} 
S_L^{-1}(p) &=& i\pslash + m + {a p^2\over 2} - \lambda \Sigma_L + O(a^2)\,, 
\\ 
\Sigma_L &=& i\pslash \Sigma_{1L} + m \Sigma_{2L} 
+ a p^2 \Sigma_{3L} +  a i\pslash m \Sigma_{4L} + a m^2 \Sigma_{5L} 
+ {i\pslash m^2 \over p^2} \Sigma_{6L}\,, 
\\ 
\lambda &=& {g_0^2 C_F\over 16 \pi^2} \,, 
\end{eqnarray} 
where explicit results for the $\Sigma_{jL}$ 
(for arbitrary $m^2/p^2$) can be deduced from 
the results in Ref.~\cite{capitani}. 
Here the quark mass is defined as usual 
\begin{equation} 
am = {1\over 2\kappa} - {1\over 2\kappa_c} \,, 
\end{equation} 
using the one-loop result for $\kappa_c$. 
We need only the large $p$ behavior of the $\Sigma_{jL}$, 
\begin{equation} 
\Sigma_{jL} = \ell_j \ln( a^2 p^2 ) + d_j + O(m^2/p^2) \,. 
\end{equation} 
The coefficients of the logarithm are (using the 
appropriate leading order value, $c_{SW}=1$) 
\begin{equation} 
\ell_1 = \ell_3 = \alpha\,,\quad 
\ell_2 = - \ell_4 = -2 \ell_5 = (3 + \alpha)\,,\quad 
\ell_6 = 0 \,, 
\label{eq:logparts} 
\end{equation} 
where $\alpha$ is the gauge parameter which vanishes in Landau gauge. 
It is important to note that subleading terms proportional 
to $(m^2/p^2)$ are absent from $\Sigma_{1L}$ and $\Sigma_{3L}$, 
since these are explicitly included in 
$\Sigma_{6L}$ and $\Sigma_{5L}$, respectively. 
These terms do contribute in the large $p^2$ limit. 
However, the other such terms (in $\Sigma_{2L}$ and $\Sigma_{4L}-\Sigma_{6L}$) 
give vanishing contributions in this limit.\footnote{%
We thank Paul Rakow for pointing out to us the importance of including the 
apparently subleading form factor $\Sigma_{6L}$, 
as has been done in Ref.~\cite{capitani}.} 
 
Using these results, we can evaluate the r.h.s. of Eq.~(\ref{eq:impSinv}) 
\begin{eqnarray} 
&&{\hat{S}^{-1}(p) \over Z_q^0} = i\pslash (1 - \lambda \Sigma_{1L}) 
+ m (1 - \lambda \Sigma_{2L}) 
\nonumber\\ 
&&\mbox{} + a p^2 
\left[2 \cp_q (1 - 2\lambda \Sigma_{1L}) 
     +2 c_{NGI} (1 - \lambda \Sigma_{1L}) 
     + 1/2 - \lambda \Sigma_{3L} \right] 
\nonumber \\ 
&&\mbox{} + a i\pslash m 
\left[-4 \cp_q (1 - \lambda \Sigma_{1L}-\lambda \Sigma_{2L}) 
     -2 c_{NGI} (1 - \lambda \Sigma_{2L}) 
     +b_q(1 -\lambda \Sigma_{1L}) 
     - \lambda \Sigma_{4L} \right] 
\label{eq:pertSinv} \\ 
&&\mbox{} + a m^2 
\left[- 2 \cp_q (1 - 2\lambda \Sigma_{2L} + 2 \lambda \Sigma_{6L}) 
      + b_q (1 -\lambda \Sigma_{2L}) 
      - c_{NGI} 2 \lambda \Sigma_{6L} 
      - \lambda \Sigma_{5L} \right] + O(a^2,\lambda^2) \,. 
\nonumber 
\end{eqnarray} 
    From this we can deduce the constraints on the improvement coefficients. 
First we note that, for large momenta, 
\begin{equation} 
S^{-1}(p) = \hat{S}^{-1}(p) + O(a^2) = -{1 \over p^2 \Sigma_1(p^2)} 
\left(i\pslash -{\Sigma_2(p^2)\over\Sigma_1(p^2)} \right)+O(m_q^2/p^2)\, , 
\end{equation} 
where we have used Eqs.~(\ref{eq:impcond}) and~(\ref{eq:decomposition}). 
Asymptotically, $\Sigma_1(p^2)$ is independent of $m_q$, 
while $\Sigma_2/\Sigma_1 = m_q$ up to logarithmic corrections in $p^2$. 
It follows that, asymptotically, there are no terms proportional to 
$p^2$ and $i\pslash m$ in $\hat{S}^{-1}$, and so the coefficients 
of these terms in Eq.~(\ref{eq:pertSinv}) must vanish. 
These constraints can be written 
(up to corrections of $O(\lambda^2)$) 
\begin{eqnarray} 
2\cp_q+ 2c_{NGI} + {1/2} &=& \lambda\left[ 
(4\cp_q + 2 c_{NGI} + 1)\Sigma_{1L} + (\Sigma_{3L}-\Sigma_{1L}) \right] 
\,, 
\label{eq:ppconstraint}\\ 
4\cp_q+ 2c_{NGI} - b_q &=& \lambda\left[ 
(4\cp_q -b_q)\Sigma_{1L} + 
(4\cp_q +2 c_{NGI} + 1)\Sigma_{2L} - 
(\Sigma_{2L}+\Sigma_{4L}) \right] 
\,. 
\label{eq:mpconstraint} 
\end{eqnarray} 
 
The final constraint is obtained by noting that 
$\hat{S}^{-1}$ depends linearly on the renormalized mass $m_q$, 
which is related to the lattice quark mass by 
\begin{equation} 
m_q = Z_m m (1 +a b_m m) \,. 
\end{equation} 
It follows that the ratio of $m^2$ and $m$ terms in 
Eq.~(\ref{eq:pertSinv}) gives $b_m$. 
In this way one finds 
\begin{equation} 
b_m + 2\cp_q - b_q = \lambda\left[ 
(2\cp_q + 1/2) \Sigma_{2L} 
- (4\cp_q+ 2 c_{NGI}) \Sigma_{6L} 
- (\Sigma_{2L}/2+\Sigma_{5L}) \right] 
+O(\lambda^2) \,. 
\label{eq:mmconstraint} 
\end{equation} 
 
The three equations (\ref{eq:ppconstraint}), 
(\ref{eq:mpconstraint}) and (\ref{eq:mmconstraint}) 
provide non-trivial constraints because their l.h.s.'s are 
constants, independent of $ap$, while the r.h.s.'s contain 
terms with logarithmic dependence on $ap$. 
At first sight, the system is over-constrained, because each 
of the three equations has an $O(1)$, an $O(g_0^2)$, 
and $O(g_0^2 \ln(ap))$ part, providing nine constraints in total, 
while there are four improvement constants, each with an $O(1)$ 
and an $O(g_0^2)$ part to be determined. 
It turns out, however, that the constraints are redundant, and, 
in fact, the system is under-constrained. 
 
To proceed we expand the improvement constants in powers of $\lambda$, 
e.g.\footnote{%
Note that it is standard to expand in powers of $g_0^2$ rather 
than $\lambda$. We choose to expand in $\lambda$ so as to simplify equations.} 
\begin{equation} 
b_q = b_q^{(0)} + \lambda b_q^{(1)} + O(\lambda^2) \,. 
\end{equation} 
Then the first constraints are that the l.h.s.'s of the three equations 
vanish at leading order, since the r.h.s.'s are of $O(\lambda)$. 
This gives 
\begin{equation} 
2{\cp_q}^{(0)}+ 2c_{NGI}^{(0)} + {1/2} = 0 \,,\quad 
4{\cp_q}^{(0)}+ 2c_{NGI}^{(0)} - b_q^{(0)} = 0 \,,\quad 
b_m^{(0)} + 2{\cp_q}^{(0)} - b_q^{(0)} =0 \,. 
\end{equation} 
We see that, as noted in the text, we cannot determine the tree-level 
value of all four constants with a tree-level computation of the 
propagator. Only the combinations ${\cp_q}^{(0)}+ c_{NGI}^{(0)}=-1/4$, 
$2{\cp_q}^{(0)} - b_q^{(0)}=1/2$ and $b_m^{(0)}=-1/2$ are determined. 
 
The one-loop computation does, however, completely determine the 
tree-level improvement coefficients. 
To see this, note that we have written the three equations so that 
the combination of $\Sigma_{jL}$ in the last term on each r.h.s. 
has no dependence on $\ln(a^2p^2)$ [see Eq.~(\ref{eq:logparts})], 
and is thus just a constant (in any covariant gauge). 
In Landau gauge, $\Sigma_{1L}$ is also a constant, but $\Sigma_{2L}$ 
has a logarithmic part. To cancel this logarithmic part on the 
r.h.s.'s of Eqs.~(\ref{eq:mpconstraint}) and~(\ref{eq:mmconstraint}) 
requires 
\begin{equation} 
4{\cp_q}^{(0)} +2 c_{NGI}^{(0)} + 1 = 0 \,,\qquad 
2 {\cp_q}^{(0)} + 1/2 = 0 \,. 
\label{eq:const2} 
\end{equation} 
These two extra equations could be inconsistent with the previous three, 
and is a non-trivial test of improvement that they are not. 
They provide the extra constraint needed to determine the 
tree level improvement coefficients, and one finds that 
\begin{equation} 
c_{NGI}^{(0)} = 0 \,. 
\end{equation} 
Thus the ``new'' constant does vanish at tree level, but from the 
point of view of the calculation this is an accident. For the 
tree-level values of the other two constants one finds 
${\cp_q}^{(0)}=-1/4$ and $b_q^{(0)}=1$. 
 
We note in passing that the coefficient of $\Sigma_{1L}$ 
in Eq.~(\ref{eq:ppconstraint}) does vanish due to Eq.~(\ref{eq:const2}). 
This means that the improvement works in perturbation theory 
at one-loop in any covariant gauge. 
 
Finally, we can deduce the constraints from the non-logarithmic 
$O(\lambda)$ terms. These can be written 
\begin{eqnarray} 
2{\cp_q}^{(1)}+ 2c_{NGI}^{(1)} &=& \Sigma_{3L}-\Sigma_{1L} 
\,,\\ 
2{\cp_q}^{(1)} - b_q^{(1)} &=& 
-\Sigma_{3L}+\Sigma_{1L} - \Sigma_{2L}-\Sigma_{4L} 
 \,,\\ 
b_m^{(1)} &=& 
\Sigma_{3L}-\Sigma_{1L} + \Sigma_{2L}/2 +\Sigma_{4L} - \Sigma_{5L} 
+ \Sigma_{6L} 
 \,. 
\end{eqnarray} 
Thus a one-loop computation of the propagator does not 
determine $c_{NGI}^{(1)}$. To do so would require a two-loop 
computation of the propagator. Alternatively, noting that 
the gauge non-invariant term can be rewritten as $\Aslash q$, 
it seems likely that a one-loop computation of the quark-gluon vertex would 
determine $c_{NGI}^{(1)}$. 
 
\section*{Acknowledgements} 
This work was partially supported by EU grants CHRX-CT92-0051 and 
HTRN-CT-2000-00145. M. Talevi acknowledges the hospitality of the 
University of Rome ``La Sapienza'' where most of the work was 
carried out, the INFN for partial support and EPSRC for its 
support through grant GR/K41663. C.T.S. acknowledges PPARC for its 
support through grants PPA/G/S/1997/00191 and PPA/G/S/1998/00525. 
G.M., G.C.R., M.T. and M.T. acknowledge partial support by 
M.U.R.S.T. S.S. thanks the University of Rome ``La Sapienza'' for 
its hospitality and INFN for partial support. S.S. was also 
partially supported by the U.S. Department of Energy through grant 
DE-FG03-96ER40956.

\end{document}